\documentstyle[12pt]{article}

\begin{document}

\baselineskip=18.6pt plus 0.2pt minus 0.1pt


\makeatletter
\@addtoreset{equation}{section}
\renewcommand{\theequation}{\thesection.\arabic{equation}}
\begin{titlepage}
\title{
\hfill\parbox{4cm} {\normalsize UFR-HEP/01-01}\\ \vspace{1cm}
     {\bf         Mirror Symmetry and  Landau Ginzburg
        Calabi-Yau Superpotentials in F-theory Compactifications }
}
\author{
Adil. Belhaj\thanks{{\tt ufrhep@fsr.ac.ma}}
  {}
\\[7pt]
{\it  High Energy Physics Laboratory,  Faculty of sciences,  Rabat,  Morocco}
}

\maketitle \thispagestyle{empty}
\begin{abstract}
 We study  Landau Ginzburg   (LG)  theories   mirror  to  2D
 $N=2$     gauged  linear sigma  models on
  toric  Calabi-Yau manifolds.  We  derive and  solve   new
    constraint  equations  for  Landau Ginzburg   elliptic  Calabi-Yau
     superpotentials,  depending on the physical data of  dual
       linear sigma models.   In    Calabi-Yau
         threefolds  case, we  consider  two  examples.  First,  we
          give  the mirror symmetry of the canonical  line  bundle
             over   the   Hirzebruch surfaces  $\bf F_n$.  Second, we
               find  a  special geometry  with  the   affine  so(8)
                 Lie algebra
                 toric data extending the geometry of elliptically fibered
                  K3.   This geometry   leads to  a pure $N=1$ six dimensional   SO(8) gauge model from the F-theory
                   compactification. For Calabi-Yau  fourfolds,  we   give
                    a  new    algebraic  realization  for  ADE
                      hypersurfaces.
\end{abstract}
\newpage
\tableofcontents
\newpage
\end{titlepage}
\newpage
\def\be{\begin{equation}}
\def\ee{\end{equation}}
\def\bea{\begin{eqnarray}}
\def\eea{\end{eqnarray}}
\def\nn{\nonumber}
\def\l{\lambda}
\def\t{\times}
\def\[{\bigl[}
\def\]{\bigr]}
\def\({\bigl(}
\def\){\bigr)}
\def\p{\partial}
\def\o{\over}
\def\ta{\tau}
\def\cm{\cal M}
\def\R{\bf R}
\def\b{\beta}
\def\a{\alpha}
\section{Introduction}

  One of  the  most beautiful  properties of  type II superstrings
    is  that  type  IIA   string  propagating  on  a   Calabi-Yau
       $M$  may behave identically with    type  IIB   string
        propagating  on  a different    Calabi- Yau  $W$.  In this
         way,  the complex (Kahler) structure  moduli space of  $M$ is
          identical to the Kahler  (complex) structure  moduli space of $W$.
             The pairs of manifolds satisfying this  map  are known as
              mirror pairs, and this string duality is  called mirror
               symmetry \cite{syz,fhsv,st}.   This symmetry  plays  also
                 an  important  role   in the  geometric engineering  of
                  4D $N=2$   quantum field theories (QFT), embedded
                      in type  II superstring theories  on  singular
                      Calabi-Yau
                       threefolds, where this map can be  used  to
                          obtain exact results for the type IIA superstring
                           Coulomb
                           branch
                           \cite{kkv,kmv,m,bfs1,bfs2}.  \\
 Recently, mirror symmetry  has been   used in the context of 2D
  superconformal   field  theories  with boundaries involving   $N=2$
   sigma models  (SM) and Landau -Ginzburg (LG) theories.
     It  was  shown
      in   \cite{hv,hiv,av}  how   these   models  can be  related
        by mirror symmetry.  This leads to the map  between   A-type
         branes, wrapping special  Lagrangian submanifolds, in the sigma model approach   and B-type branes wrapping
           holomorphic cycles
           in the context of  LG theories.   This  link   has been a
           powerful  tool   in   the study  of  algebraic
             realizations of  Calabi-Yau  manifolds.  In particular,
              elliptic geometries  used    in     the  derivations  of  non
               perturbatif  superstring solutions   from  either   D-brane
               physics or   F-theory
                compactifications \cite{vf,vm}.   \par The aim of
                this paper is to contribute in this program by   deriving    new  classes
                 of constraint  equations   for LG  elliptic  Calabi-Yau
superpotentials using  the recent derivation of mirror symmetry in
toric sigma model. In particular, we find a special elliptic
Calabi-Yau threefolds extending  the mirror superpotentials of the
blow-up of  the affine ADE  local $ K3$, used  in the geometric
engineering  of $4D$ $N=2$ superconformal field theories
\cite{kmv,bfs2}.  This involves the affine  so(8) Lie algebra as
Mori vectors  toric data leading to a new  $N=1$ SO(8) gauge model
in six dimensions from F-theory compactifications point of view.
\\ The organization of this paper is as follows. In section 2, we
give an overview  of Vafa's  construction of F-theory.  Then  we
study F-theory  on Calabi-Yau spaces with elliptic geometric
structures and  the role they play in the understanding  of   the
lower dimensional superstring models. In section 3,  we first
discuss aspects  of $2D$ $N=2$    linear sigma model.  Second  we
study the interplay between this model and  toric geometry  which
plays a crucial role  for us later in this paper. We also
introduce mirror symmetry, as  made in \cite{hv,hiv,av},  to
obtain  the LG mirror   theory. Then  we illustrate  the example
of elliptic $K3$ with  ADE singularities, to engineer $N=1$  gauge
theories in eight dimensions with ADE gauge groups from F-theory
compactifications. In section 4, we  consider two examples of  the
mirror symmetry for sigma model on Calabi-Yau threefolds. First,
we give the mirror theory of the linear sigma model on the
canonical line bundle over the Hirzebruch surfaces $\bf F_n$. This
geometry recovers  the leading example of $\bf F_0$  studied in
the context of   the mirror action of Lagrangian D-branes
\cite{av}.   Second we  find  a special elliptic   and $K3$
fibered  Calabi-Yau threefolds extending the elliptic  ADE mirror
superpotentials   to Calabi-Yau threefolds with affine so(8)   Lie
algebra toric data. This background space gives, from F-theory
compactifications,  a new pure $N=1$ SO(8) Yang-Mills theory  in
six dimensions.   In section 5, we use the techniques developed in
section 3  to derive a solution for the mirror superpotentials
associated to ADE Calabi-Yau fourfolds hypersurfaces.   This gives
a  toric realization  of ADE Calabi-Yau fourfolds hypersurfaces
studied in \cite{gvw} in the context of derivations of  $2D$
superconformal field theories from superstring compactifications.
In section 6 we give our conclusion.

\section{ Generalities  on   F-theory}
\subsection{  Review on Vafa's construction of F-theory}
  F-theory defines a non perturbative vacua of type IIB superstring
  theory in which the dilaton $(\phi)$  and  the axion $(\chi)$ fields of
   the superstring are not constants. These fields  are  known as the
complex string  coupling  moduli  $\tau_{IIB}=\chi +ie^{-\phi}$
which is interpreted as the complex parameter of an elliptic curve
leading then to non pertubative  vacua of type IIB superstring
        theory  in a twelve dimensional spacetime \cite{vf}.  F-theory
         may be  also defined by help of superstring  dualities.
           As we will see later on,  F-theory on elliptically fibered
             Calabi-Yau  spaces  may also be defined in terms of dual
              superstring models, but let us first review briefly some
              features of this theory.  Type IIB  is
               a ten dimensional theory of closed superstrings with chiral
                $N=2$ supersymmetry.  The bosonic fields of the
                 corresponding low energy field theory are the
                  graviton $g_{\mu\nu}$, the antisymmetric tensor
                   $B_{\mu\nu}$ and the dilaton $\phi$ coming
                    from the NS-NS sector and the axion $\chi$,
                     the antisymmetric tensor fields $\tilde B_{\mu\nu}$
                      and the self dual four form $D_{\mu\nu\sigma\lambda}$
                       coming from R-R sector.  As we see,  there
                        is no non abelian gauge field in the massless
                         spectrum of type IIB superstring theory
                           but instead contains   D$p$-branes  solitonic
                            objects, with $p=-1,1,3,5,7$ and  $9$, on which
                             live $A_{\mu}$ gauge  fields of open string
                              field  theory  \cite{p}.  These
                               extended objects are non perturbative
                                solutions playing a crucial role in
                                 string dualities,  and  in  the embedding
                                   of  QFT's  in superstring models  by
                                   using either Hanany-Witten method
                                      \cite{hw,wbrane}  or  geometric
                                      engineering approach
                                         \cite{kkv,kmv,m,bfs1,bfs2,b}.
                                         Type IIB  superstring theory has a
                                          non perturbative SL(2,Z) symmetry
                                           for which the fields $g_{\mu\nu}$
                                            and $D_{\mu\nu\sigma\lambda}$ are
                                             invariant but the complex
                                              string coupling
                                               $\tau_{IIB}=\chi +ie^{-\phi}$
                                                and the doublet $(B_{\mu\nu},\tilde B_{\mu\nu})
                                                $ of two forms are believed to transform as
                                                 \cite{sch} :
\be
\tau_{IIB} \to {a\tau_{IIB}+b\over c\tau_{IIB}+d}, \quad  a,b,c,d
\in {\bf Z}, \ee and
\be
{B_{\mu\nu}\choose \tilde B_{\mu\nu}}= \left(\matrix{
a&b\cr
c&d\cr}\right){B_{\mu\nu}\choose \tilde B_{\mu\nu}},
\ee
 where the integers $a,b,c$  and  $d$ are such that $ ad-bc=1$.\\
Following Vafa \cite{vf}, one may interpret the complex field $\tau_{IIB}$
as the complex structure  $\tau_{ T^2}$  of an extra torus $T^2$:
\be
\tau_{IIB}=\chi +ie^{-\phi}=\tau_{T^2}.
\ee
 This  extra torus $T^2$ combines  with the ten spacetime dimensions
   to give  a  twelve dimensional theory.  From this view,  $10D$ type
    IIB superstring  theory may be seen as the compactification
     of F-theory on the elliptic curve $T^2$:
\be
 \textrm{Type  IIB superstring theory }
  \sim {  \textrm{ F-theory} \o T^2} .
\ee
\subsection{F-theory compactifications  and string dualities}
 Here we study the F-theory compactifications and their connections to string models. To do
 so, we consider first a
  $(n+1)$-dimensional  Calabi-Yau  manifold  $ W_{n+1}$
which  has an elliptic fibration over a $n$-dimensional complex
base space $B_{n}$
\be
 y^2=x^3+f(z_i)x+g(z_i),\quad z_i \in B_{n},
\ee
 where $z_i$  are   the  local coordinates of   $B_{n}$.   F-theory
   compactification on $W_{n+1}$ is equivalent to type IIB  superstring
    theory  on $B_n$ with  the  varying complex string
     coupling  $\tau_{IIB}$:
\be
\chi (z_i) +ie^{-\phi (z_i)}=\tau _{T^2}(z_i).
\ee
 The positions  of the degenerate elliptic fibers on  $B_n$  are  given by
   the solution of the following  equation
\be
\delta= 27g^2(z_i)+4f^3(z_i)=0, \ee
 where $\delta$ is the discriminant of  the elliptic fibration.
 Recall that the well  known example of    F-theory compactification
    is  the eight dimensional  model \cite{vf,vm}.  This   is obtained  by
      the compactification on  elliptically fibered  $K3$ surface
\be
 y^2=x^3+f(z)x+g(z),\quad  z \in {\bf P^1}
\ee
 In this case,  the functions  $f$ and $g$  are polynomials of degree
  8 and 12 in $z$ respectively:
\bea
f(z)=\sum\limits_{i=0}^8a_iz^i\nn\\
g(z)=\sum\limits_{i=0}^{12}b_iz^i.
\eea
 The   complex structure $\tau_{T^2}$  is  now  a  function    of  one
   variable  $z$   (local coordinate of   $ \bf P^1$ )  which  varies over
    the $ \bf P ^1$ base of elliptically fibered  $K3$.  Equation (2.7)
    has generically 24 singular points corresponding   to  the neutrality
     condition  of the discriminant $\delta$.  These singularities have a remarkable
      physical interpretation. To each one of the 24 points,
        it is associated to  the location of  a  D7-brane of non perturbative
type IIB superstring theory.\\
\\
 {\bf  F-theory /heterotic
duality }\\
 F-theory on elliptically fibered  $K3$   leads  to new  type    IIB
   superstring  theory solutions in eight dimensions.  This model  is conjectured to be dual
     to  the heterotic superstring theory on $T^2$,
\be
 {  \textrm{ F-theory} \o K3} \sim   { \textrm{ heterotic  superstring  }\o T^2} ,
\ee
 with the heterotic string  coupling constant  $g_s^h$  is given by the size of the $\bf P^1$
  base of elliptically fibered $K3$.  This eight dimensional model
   can be further compactified
  to lower dimensions by fibering both sides over the same
  complex base $ B_{n-1}$ using   the so-called adiabatic principle \cite{vw}.  In this way,  the above eight dimensional duality becomes \cite{bm,m1,m2}
\be
 {  \textrm{ F-theory} \o {W_{n+1}}} \sim   { \textrm{ heterotic  superstring  }\o Z_n} ,
\ee
 where  $W_{n+1}$  has  $K3$  fibration over $ B_{n-1}$  (with   $K3$=$W_2$).  It also has an elliptic fibration,   inherited from the elliptic
    fibration of  $W_2$, over  $B_{n}$ .  While the
          heterotic Calabi-Yau manifold $Z_n$ is   an elliptic fibration
           over the  base  $B_{n-1}$.    For example,   if we  fiber
              eight dimensional data  over   an    extra  torus
               $ B_1=T^2$,   then the resulting duality becomes
                a duality between F-theory on K3$\t T^2$ and heterotic
                string on $T^4$.   The latter is known to be dual to type
                 IIA string on  K3 \cite{ht,wd,ketal,kv,asp}.
                 Thus,
                   interesting  superstring model in lower dimensions  can
                      be obtained   from  F-theory compactifications on
                       elliptically fibered  Calabi-Yau
                      manifolds.  This
                         gives a pure geometric interpretation of the
                          perturbative  superstring spectrum and determines
                           at the same time the non perturbative dynamics
                             associated to D-brane physics  in type II
                              superstring theories or   to  singular bundle
                               of   $N=1$ superstring models.
                                  In what follows we  shall use the   sigma
                                   model/LG  mirror correspondence  to
                                      develop  new
                                    algebraic realization of elliptic Calabi-Yau manifolds
                                     involving both elliptic fibration and  $K3$ fibration.
                                       Special attention will be given to LG elliptic
                                       Calabi-Yau 3-4 folds
                                       superpotentials.
\section{Mirror symmetry in   $2D$ $N=2$ field theory}
  \subsection{ $2D$ $N=2$ Sigma model}
In this section  we study  the gauged linear sigma model
introduced by Witten  as  a field theoretic  description of
Calabi-Yan manifolds \cite{w}. Then  we discuss the corresponding
LG mirror
    theory  studied    in \cite{hv,hiv,av}.
     For simplicity,  we consider an abelian   gauge group
      $U(1)^r$  described  by   superfields $V_a$ $( a=1,\ldots,r)$. We assume that there are
        $k$ charged  chiral   superfields  $\Phi_i$ $( i=1,\ldots,k) $   of  vector charges
         $q_i^a$  $ (a=1,\ldots,r)$ \cite{w,agm}. The Lagrangian of this model,  in terms
          of superfields language, reads as
\be
{\bf L}= \int d^2x d^4\theta \sum\limits_{i=1}^k {\bar \Phi_i}
e^{2{q^a_i}V_a}\Phi_i-\sum\limits_a\rho_a \int d^2x  d^4\theta
V_a+( \int d^2x d^2\theta
 W(\Phi)+hc).
 \ee
Integrating  with respect to $\theta$,  we find   the
superpotential  energy for
 the dynamical  scalar  fields $\phi_i$
\be
U(\phi)=\sum \limits _{i=1}^k|{\p W\o \p \phi_i}|^2+\sum \limits
_{a=1}^r {1\o 2e^2_a} D_a^2,
\ee
 where  $W$ is the superpotential and  $e_a$'s  are  the gauge  coupling parameters.
   $ D^a$ are known as   D-terms:
\be
D_a^2=\sum\limits _{a=1}^{r}e^2_i (\sum \limits _{i=1}^{k}q_i^a
|\phi_i|^2-\rho_a)^2, \ee where the $\rho_a$ parameters are
Fayet-Illiopoulos (FI) terms with  the  $\theta$  angles  give
complexified  Kahler parameters:
\be
   t_a= \rho_a +i\theta_a,\quad a=1,\ldots,r.
\ee
 We next    suppose  that  there is no superpotential  for the charged matters
$$ W(\phi)=0.$$
 Thus,  the minimum of the potential energy comes only  from the   D-terms.
 Vanishing of these terms give us
\be
\sum \limits _{i=1}^{k} q_i^a |\phi_i|^2=\rho_a.
\ee
 Dividing the space of solutions  ( which will be called  $M$) of (3.5) by
 the  complexified  gauge group   $U(1)^r$
\be
\phi_i\to e ^{iq^a_i\gamma_a}\phi_i, \ee
 we find the following  complex space
\be
{C^{k}\over {C^*}^r}, \ee where $ C^k$ corresponds to   complex
coordinates  $z_i$  and   the  $ C^*$ actions are  given by
\be
z_i   \to \lambda ^{q_i^a} z_i ,\quad  i=1,2,\ldots ,k ; \;\;
 a=1, 2,\ldots ,r.
\ee
 For example, if we have a $U(1)$ gauge theory  with two chiral fields with charges 1,
 the classical moduli space is  $\bf P^1$.\\
  The  space of solutions  ($M$)   we have been describing  has
a nice geometrical interpretation in terms of toric geometry
\footnote{ For more details on toric geometry, see
   \cite{f,c,  lv,cpr,r,bs3}.}.  This  has been a  beautiful  interplay
      between  $2D$ $ N=2$ sigma models and toric geometry \cite{w,kmv}.
 Indeed   interpreting the previous   $(\phi_i)$ matter fields as the  $z_i$
  coordinates  of  ( 3.7) and the $ q_i^a $  quantum charges,
    under  the  $ U(1)^r$ symmetry, should be interpreted as the  Mori
     vectors  of  toric geometry language.  In this way,
     the vacuum space may have a  toric diagram  ${\Delta}$ which
      consists of   $k$ vertices  $\{{v_i}\}$  in  the standard lattice
       $ \bf Z^n$, where $ n=k-r$   is   the complex dimension of
        the  space of solutions ( we are assuming  that there is
         no toric fibration structure).  Every vertex   $ v_i$ corresponds to a matter field
          $(\phi_i)$  in  our  $2D$ $ N=2$ sigma model.  Since the complex dimension of
           vacuum space is $n$,  there  are $r$ relations between the $k$ vertices  which
            read as
 \be
\sum \limits _{i=1}^{k} q_i^a v_ i=0, \quad  a=1,\ldots,r. \ee In
this representation, it is  worthwhile to mention the four
following: \\ 1- If the  $ q_i^a $'s are all positive definite, or
negative definite,
 the  space of solutions is compact.   However, if there is a mixture of positive and negative $ q_i^a $'s, the toric target space is non compact .\\
2-    For   $ q_i^a $'s obeying the neutrality condition
 \be
\sum \limits _{i=1}^{k} q_i^a =0, \quad  a=1,\ldots,r,
\ee
  the  toric target space is a  non compact   Calabi-Yau manifold   and
   the field  theory flows in the infrared to a non trivial superconformal model
     \cite{w,kmv,bs1,bs2}.  This type of   manifolds  plays
  a crucial role  in the study    of non perturbative  superstring theory
   compactifications,
   in particular, in  the  geometric engineering of QFT's.\\
3- If all $\rho_a$'s are zero, then the toric manifold is
singular.\\ 4- For all $\rho_a$'s $\neq 0$, we have a smooth toric
manifold.  In this
 case  the  (FI) parameters,  which  are  given  by  the  Hodge number
   $h^{1,1}(M)$ or equivalently by  the number of  $U(1)$ factors,
     are interpreted as blow up parameters of the singularity.
     \subsection{ LG mirror theory}
 Having  introduced the linear sigma model construction,
    we  will   now  discuss  the  corresponding    mirror   theory.
      There are different ways one follow  to obtain the mirror theory.
        The latter is  a   LG   model with  Calabi-Yau  superpotentials,
          depending on the number   of  chiral  multiples and gauge fields
           of dual theories.  A tricky way,  to   write down the equation of  LG  mirror
            superpotential  (dual to the previous Sigma model  ( 3.5)),  is to  introduce
              in the game      $k$ dual chiral fields $Y_i$
                to each field in the sigma model such that \cite{hv,hiv,av}
\be
 \textrm{Re}\; Y_i= |\phi_i|^2,\quad  i=1,\ldots,k.
\ee
 For convenience,  we define  new variables  $x_i$
\be
x_i=e^{-Y_i}.
 \ee
The  defining equation of the LG mirror  superpotential  takes the
form
\be
 \sum \limits _{i=1}^k x_i=0,
\ee
  where the  fields  $x_i$  must satisfy
\be
\prod \limits _{i=1}^k  x_i^{q^a_i}=e^{-t_a},  a=1,\ldots,r.  \ee
 Recall that the  $t_a$ are  the complexified  Kahler parameters of  sigma models which
  now define the complex    structure  of  LG mirror geometry.  The solution   of  these
   equations  often described by $ (n-2)$-dimensional hypersurfaces. This is not a problem
     since one can restore the correct dimension by introducing  two auxiliary  fields
       $u$ and $v$ and  equation (3.13)  becomes
\be
 W(x_i)=\sum\limits _{i=1}^k  x_i=uv
\ee
 Note that the quadratic term $uv$ does not affect the complex structure of the mirror
   superpotential.
\subsection{ Elliptic ADE  mirror superpotentials}
 As our first  example we consider  the LG   $K3$ superpotentials  with
  deformed  elliptic ADE
  singularities.     We start by  constructing   these geometries,
   with ADE singularities,    as gauged   $N=2$  two dimensional linear
    sigma model. In general,   these   are described  by
      a  $U(1)^{r+1}$  gauge group with  $(r+5)$   chiral
         multiples   $\phi_i$  with  $q^a_i$  vector charges specified
           latter on.  The  elliptic ADE spaces  of classical vacua,
             in  the absence of  the  sigma model superpotential,
              are  given by
  \be
U=\sum\limits _{a=0}^{r}e^2_a (\sum \limits _{i=1}^{r+5}q_i^a
|\phi_i|^2-\rho_a)^2,  \quad a=0,1\ldots,r, \ee where $r$ is the
rank of ADE algebras and  the $q^a_i$'s  are the quantum
 charges of $\phi_i$ under the $U(1)^{r+1}$ gauge symmetry,  up to  details  are   proportional to the Cartan matrices
$K_{ai}$ of the ADE  Lie algebras  in question.  These   vectors
charges satisfy   the  condition $\sum\limits_{i=1}^{{r+5}}q^a_i=0
$ under which  the gauge model flow in the infrared  to 2D $N=2$
superconformal field theory. The  ADE  spaces of  classical vacua
may  be  described by a toric diagram  $\Delta$ spanned by $(r+5)$
vertices
 \be
v_i=(n_i,m_i,s_i) \ee of the standard lattice $\bf Z^3$, where the
first  entry  $ n_i$  takes either zero or the Dynkin weighted of
the adjoint  representation of   the corresponding Lie algebra.
These $(r+5)$ vertices   fulfill the following   $( r+1)$
relations:
 \begin{equation}
\sum\limits_{i=1}^{r+5} q^a_iv_i=0,\quad  a=0,1...,r,
\end{equation}
 with   the  Calabi-Yau  condition
 \be
\sum\limits_{i=1}^{r+5}q^a_i=0.
\ee
Having introduced the toric data  of sigma model construction of local
   elliptic ADE  $K3$ surface, we will now apply the  mirror symmetry
     to get  the corresponding ADE superpotentials for  LG theory.
       To write down the algebraic  equations of  the  mirror
         geometry,  we will use the  toric data of sigma model construction.
          Indeed, we associate to each vertex $v_i$ of the toric diagram  $\Delta$
           a monomial $ x_1^{n_i} x_2 ^{m_i} x_3^{s_i}$,  where
             $ x_1$, $  x_2 $ and $ x_3$  are  LG gauge invariant fields
             \cite{bfs2,bs3}.  The superpotentials  of the mirror theory
                 associated  with  the   vertices (3.17)  are  described by
                   complex  $2D$  Calabi- Yau surfaces $
                   W_2(ADE)$:
\be
W_2(ADE)=  \sum\limits _{i=1}^{r+5} a_i
{x_1}^{n_i}{x_2}^{m_i}{x_3}^{s_i}=0, \ee where $ a_i$  are the
complex  parameters defining the complex structure of the mirror
superpotentials. Note that only a subset of $a_i $ are physical.
Recall that the fields $x_i$   may  be viewed  as gauge invariant
under the $ C^*$ action of weighted projective spaces  ${\bf
WP^3}$, in which the elliptic $K3$  is embedded \cite{kmv,
bfs2,bs3}, allowing us to give   a homogeneous  description of
elliptic ADE series for LG mirror superpotentials. This
homogeneous description takes the form
\be
W_2 (ADE)=  P_0(y,x,z)+ \sum\limits _{i}w^i P_i (y,x,z)=0, \ee
 where  $ (y,x,z,w)$  are  the  homogeneous coordinates of
     ${\bf WP^3}(3,2,1,\eta)$  and  where   $\eta$  is an integer
     depending on
       the type  of   Lie algebras \cite{bfs2,bs3}.  $ P_0$  describes an elliptic curve
           $\bf E$ in      ${\bf WP^2}$. In particular, a sextic in
           ${\bf WP^2}(3,2,1)$
\be
P_{0 }= y^2+x^3+z^6+ \mu  xyz=0,
 \ee
 with  $ \mu$   is a   complex
structure  moduli. The number
        of  monomials in $P_i$  is  equal  to  the number of  Dynkin
          labels equal  to $i$.
  Equation (3.21)  gives  the   ADE  mirror classification of
     local elliptically fibered $K3$    and plays a crucial role in
         geometric
engineering of
 $ N=1$  gauge theories  in eight dimensions.   Indeed, at singular limit of F-theory
  on $W_2 $,
    when   the  2-cycles  of    $ K3$  shrink
to zero area, we get enhanced    $N=1$ ADE gauge symmetries in
eight dimensions \cite{bm,m1,m2,v1,bs3}.  Recall that  the
resolution of ADE singular $K3$ occurring  in  F-theory
compactifications consists of  affine Dynkin diagrams of  a chain
of  2-cycles  with specific intersection numbers in agreement with
the corresponding affine Dynkin index. For illustration,  let us
present an  example concerning   the elliptic affine $A_{n-1}$
space (for $n$ even). The toric data of this geometry   has four
vertices: \bea v_0=(0,0,0), v_1=(0,2,3), v_2=(0,-1,0),
v_3=(0,0,-1), \eea
 describing the elliptic fiber $\bf E $ and  $n$
vertices,
\be
v_1=(1,2,3),v_{2i}=(1,2-i,3-i),v_{2i+1}=(1,2-i,2-i),\; i>0, \ee
introduced by the blow ups. The superpotential of the  mirror LG
model is
\be
 W_2 (A_{n-1})= (y^2+x^3+z^6+ \mu xyz)+w P_1(x,y,z)=0 \ee where
\be
 P_1(x,y,z)=a_nz^n+a_{n-2}z^{n-2}x+a_{n-3}z^{n-3}y+\ldots+ a_0x^{{n\o2}}.
\ee For completeness we give the toric vertices for  others
elliptic geometries\\
 {\it  -$\hat D_n$ Geometry }
\begin{equation}
\begin {array}{lcr}
v_0=(1,2,3), v_1=(1,1,1)\\v_{i+1}=(2,3-i,4-i),\quad
i=1,\ldots,n-3\\v_{n-1}=(1,{1\o 2}(6-\epsilon-n ),{1\o
2}(6-\epsilon-n )),\;v_{n}=(1,{1\o 2}(4+\epsilon-n ),{1\o
2}(6+\epsilon-n ))
\\ \tilde v_n=(0,0,0),\tilde v_{n+1}=(0,-1,0),\tilde
v_{n+2}=(0,0,-1),\tilde v_{n+3}=(0,2,3)\\
q^0_i=(-2,0,1,0^{n-2},0,0,0,1),\;q^1_i=(0,1,-2,1,0^{n-4},0,0,0,0,)\\
q^2_i=(1,1,-2,1,0^{n-3},-1,0,0,0),\ldots\\
q^{n-1}_i(\epsilon=0)=(0^{n-2},1,-2,0,0,0,1,0),\;
q^{n-1}_i(\epsilon=1)=(0^{n-2},1,-2,0,-2,1,2,0),\\
q^{n}_i(\epsilon=0)=(0^{n-2},1,0,-2,-2,2,1,0),\;
q^{n}_i(\epsilon=1)=(0^{n-2},1,0,-2,0,1,0,0), i=0,\ldots,n+4
 \end{array}
 \end{equation}
  where $\epsilon=0$ (1) for $n$ even (old).\\
  \\
$- \hat E_6$ ( curve  {\bf E} in $P^2$) :\\
\begin{%
equation}
 \begin {array}{lcr}
 v_0 =(1,-1,1),\; v_1=(1,-1,-1),
v_2=(2,-1,0),\;_3=(2,-1,-1)
\\ _4 =(3,-1,-1), \;v_5=(2,0,-1), \;v_6=(1,1,-1) \\ \tilde v_7=(0,0,0),\tilde
v_8=(0,2,-1),\;\tilde v_9=(0,-1,2),\;\tilde v_{10}=(0,-1,-1)\\
q^0_i=(-2,0,1,0, 0,0,0,0,0,1,0), \; q^1_i=(0,-2,0,1,
0,0,0,0,0,0,1)\\ q^2_i=(1,0,-2,0,1, 0,0,0,0,0,0)\\
q^3_i=(0,1,0,-2, 1,0,0,0,0,0,0)\\ q^4_i=(0,0,1,1,-2,1,0,-1,0,0,0)
\\ q^5_i=(0,0,0,0,1,-2,1,0,0,0,0)\\ q^6_i=(0,0,0,0,1,-2,1,0,0).\\
\end {array}
\end{equation}
 $- \hat E_7$ ( curve  {\bf E} in $WP^2(1,1,2)$) :\\
 \begin{equation}
\begin {array}{lcr}
v_0 =(1,-2,1),\; v_1 =(2,-2,-1),\; v_2=(2,-1,0),\;v_3=(3,-2,-1) \\
v_4 =(4,-2,-1),\; v_5=(3,-1,-1),\; v_6=(2,0,-1), \;v_7=(1,1,-1) \\
\tilde v_8=(0,0,0),\;\tilde v_9=(0,2,-1),\;\tilde
v_{10}=(0,0,1),\;\tilde v_{11}=(0,-2,-1)\\ q^0_i=(-2,1,0,
0,0,0,0,0,0,0,0,1),\;q^1_i=(1,-2,0,1,0, 0,0,0,0,0,0,0)\\
q^2_i=(0,0,-2,0,1, 0,0,0,0,0,1,0)\\ q^3_i=(0,1,0,-2,
1,0,0,0,0,0,0,0)\\ q^4_i=(0,0,1,1,-2,1,0,0,-1, 0,0,0)\\
q^5_i=(0,0,0,0, 1,-2,1,0,0,0,0,0)\\
q^6_i=(0,0,0,0,0,1,-2,1,0,0,0,0)\\
q^7_i=(0,0,0,0,0,0,1,-2,0,1,-2,0,1,0,0).
\end {array}
\end {equation}
 $- \hat E_8$ ( curve  {\bf E} in $WP^2(1,2,3)$ ):\\
\begin{equation}
 \begin {array}{lcr} E_8 :\\v_0 =(1,2,3),\; v_1 =(2,2,3),\;
v_2=(3,1,1), \;v_3=(3,2,3) \\ v_4 =(4,2,3),\; v_5=(5,2,3),\;
v_6=(6,2,3),\;v_7=(4,1,2),\;v_8=(2,0,1) \\ \tilde
v_9=(0,0,0),\tilde v_{10}=(0,-1,0),\tilde v_{11}=(0,0,-1),\tilde
v_{12}=(0,2,3)\\
 q^0_i=(-2,1,0, 0,0,0,0,0,0,0,0,0,1)
 q^1_i=(1,-2,0,1,0, 0,0,0,0,0,0,0,0)\\
q^2_i=(0,0,-2,0,0,0,1, 0,0,0,0,1,0)\\ q^3_i=(0,1,0,-2,
1,0,0,0,0,0,0,0,0)\\ q^4_i=(0,0,0,1,-2,1,0, 0,0,0,0,0,0)\\
q^5_i=(0,0,0,0,1,-2,1,0,0,0,0,0,0)\\
q^6_i=(0,0,1,0,0,1,-2,1,0,-1,0,0,0) \\ q^7_i=(0,0,0,
0,0,0,1,-2,1,0,0,0,0)\\ q^8_i=(0^7,-2,0,1,0,0).
\end {array}
\end {equation}
-{\bf  BCFG models} \\ We conclude this section by  noting  that
the above  analysis  is also  valid   for
 non  simply  laced    BCFG   LG   mirror  geometries   of
    $K3$.  This  is  based  on   toric realizations  of  the standard
techniques of the folding  of  the Dynkin nodes of  ADE  graphs
which are permuted by  outer-automorphism groups \cite{bfs2,bs3}.
Indeed, starting form the  toric  data of  the ADE simply laced
geometries considered above  one gets the constraint  equations of
the folding of  toric vertices of non simply laced  geometries
using the well known  results:
\begin{equation}
\begin{array}{lcr}
 D_{n+1}/Z_2&\to B_n\\
A_{2n-1}/Z_2&\to C_n\\
E_6/Z_2&\to F_4\\
D_4/Z_2 &\to G_2.
\end{array}
\end {equation}
\section{LG Calabi-Yau threefolds superpotentials }
 In this section we  will   describe  the  LG  theory mirror
  to sigma model on
 toric  Calabi-Yau  threefolds and the role they play in the
   description of F-theory  vacua in six dimensions.  In
   particular,
      supersymmetric   QFT's   limit  of  low effective models of
         F-theory  on  singular   Calabi-Yau threefolds.
             We do not attempt to give a classification,  but instead  we
               will consider  two examples.   We  first study the
                 LG theory mirror to linear  sigma model on the canonical
                  line  bundle over the Hirzebruch surfaces $\bf F_n$.
                      This   geometry recovers   the leading example of  $\bf F_0$
                        studied in the context of   the mirror action
                         of Lagrangian  D branes  \cite{av}.  We will  see  that the mirror
                          geometry  has  also  an
                           elliptic fibration structure
\be
f(x_1,x_2)=uv.
\ee
 where  $ f(x_1,x_2) =0$  describes a  Riemann surface, $  x_1,x_2$  are  $ C^*$
  coordinates  and  $u,v$ are $\ C$ coordinates.    Second  we will consider a special
    mirror geometry  extending the $K3$ mirror superpotantials with ADE
    singularities studied in section 3.
\subsection{Elliptic fibration models}
 Let us start with the first example describing the sigma model on
   the  canonical line bundle over the Hirzebruch surfaces
    $\bf F_n,\;(n\geq 0)$ .  Recall  by the way that
      the $\bf F_n$ geometries are  defined by a non-trivial fibration of a $\bf P^1$  fiber
          on a $\bf P^1$ base.   These geometries are realized as the vacuum manifold of the
         $ U(1) \times U(1)$ gauge theory with four chiral fields with charges
         \bea
          q_i^{(1)}&=&(1,1,0,-n)\nn\\
         q_i^{(2)}&=&(0,0,1,1).\eea
These surfaces have  a  nice realization  in terms of
           toric geometry techniques \cite{f}.  This represented by four vertices
             in $\bf Z^2$ as follows
\bea
{ v_1}&=&(1,0)\nn\\
{ v_2}&=&(-1,n)\nn\\
{ v_3}&=&(0,1)\\
{ v_4}&=&(0,-1)\nn.
\eea
These vertices satisfy the following  linear toric relations
\bea
{\ v_1}+{ v_2}+n{ v_4}=0\nn\\
{\ v_3}+{ v_4}=0.
\eea
  Note that   the $\bf F_n$ surfaces are not Ricci-flat. However  they
    can be viewed  as  a part of a local geometry of a Calabi-Yau
     manifold, where there are extra dimensions.   In particular,
       if  we embed  these surfaces  in a  Calabi-Yau 3-folds
         there is a normal direction corresponding  to  line  bundle
          on  $\bf F_n$.   The  Calabi-Yau condition  requires  that the
           normal bundle must be  a canonical line bundle.
             Thus the canonical line  bundle over $\bf F_n$
               are   local  Calabi-Yau  threefolds \footnote{ Recall
                  that  for the leading example  corresponding
                    to  $\bf F_0=   \bf P^1\t  \bf P^1$ ( trivial
                     fibration),  the canonical line bundle of  $\bf F_0$
                     looks like as  the  $A_1$  ALE  space
                      ( local $K3$ surface)   fibered  over a $\bf P^1$ base
                       space \cite{kmv}.}.    These   geometries  are  used
                          in  superstring theory compactifications,  in
                           particular,   in   the  geometric engineering
                            of   $ 4D$   $ N=2$ supersymmetric gauge
                            theories, where  these  background
                            spaces allow  us to  rederive  the Seiberg-Witten   models
                                \cite{swg1,swg2}.  Roughly speaking,  the     canonical line bundle of  $\bf F_ n$
                                  surfaces  is  described by    a   $U(1)\t U(1)$   linear
                                   sigma model  with  five  matter  fields  $\phi_i$   with
                                       two  vector charges
\bea q_i^{(1)}&=&(1,1,0,-n,n-2)\nn\\ q_i^{(2)}&=&(0,0,1,1,-2).
\eea The D-flatness conditions  of  this model read  as \bea
|\phi_1|^2+|\phi_2|^2-n|\phi_4|^2+( n-2)|\phi_5|^2&=&\rho_1\nn\\
|\phi_3|^2+|\phi_4|^2-2|\phi_5|^2&=&\rho_2. \eea This classical
vacuum  has a   geometrical  realization in terms of the following
toric data
\be
\sum\limits_{i=1}^5q_i^a{ v_i}=0,\quad \sum\limits_{i=1}^5q_i^a=0,
\ee
 where the vertices    $  v_i$, which  are elements  of the  standard lattice  $\bf
 Z^3$, are  given by
\bea { v_1}=(1,0,1),\; { v_2}=(-1,n,1),\; { v_3}= (0,1,1),\; {
v_4}=(0,-1,1),\; { v_5}=(0,0,1), \eea
 and    $\sum\limits_{i=1}^5q_i^a=0$  is  the  Calabi-Yau condition
  to ensure the cancellation of the first  Chern class   $c_1=0$.
  Using equations (3.13-14), the  LG   mirror   superpotential  is obtained   by  solving the
   following constraint equations
\bea W_3(x_i)&=&x_1+x_2+x_3+x_4+x_5,\\ x_1x_2&=&e^{-t_1} x_4^n
x_5^{2-n}\\ x_3x_4&=&e^{-t_2}x_5^2. \eea After a direct
computation in the patch  $x_5=1$,   we get
\be
 f_n(x_1, x_4)=1+x_1+{e^{-t_1} x_4^n\o x_1}+x_4+{e^{-t_2}\o
 x_4}=0.
\ee
 This  LG mirror geometry  has  naively 1-dimensional Riemman surface.
  This not a problem since the LG  mirror    superpotential
  encodes all the  informations of  sigma model physical  Kahler parameters;  and
      one can restore the correct dimension,
        which is 3, by introducing the irrelevant quadratic
         term $uv $ in equation (4.9).  Thus,  the defining equation
          for  the   LG  mirror  superpotential becomes
\bea W_3(x_i,u,v)&=&f_n(x_1,x_4)-uv=0\nn\\ &=&1+x_1+{e^{-t_1}
x_4^n\o x_1}+x_4+{e^{-t_2}\o
 x_4}-uv=0, \eea which   now describes
a non compact toric  Calabi-Yau  3-folds, moreover,  permits  us
to  go beyond  the $\bf F_0$  case used in  \cite{av}. Equation
(4.12) implies    that  this  geometry  has an elliptic fibration
model over $ \bf C^2$ with coordinates $u$ et $v$  whose the fiber
is a Riemann surface  \be f_n(x_1,x_4)=0; \ee
  where   for   $n=0$ we have  an   elliptic curve in   the   ${\bf P^2}
 $   projective  space.  To see this,  we shall proceed in two steps as
  follows.  First, we consider the LG  fields $x_1$ and $x_4$
   as two invariant gauge fields  under $C^*$ action of the  ${\bf P^2}$ in
     which the  fiber is embedded:
\bea
x_1&=&{x\o z}\nn\\
 x_4&=&{y\o z},
\eea where  $x$, $y$ and $z$
 are the homogeneous variables of the  ${\bf P}^2$
 $$(x,y,z)\to (\lambda x,\lambda y,\lambda z).$$
  Second,  putting  the   equation  (4.15) in (4.13) for $n=0$
  \be
f _0(x_1,x_2)=1+x_1+{e^{-t_1} \o x_1}+x_4+{e^{-t_2}\o x_4}; \ee
 and  multiplying  by  $xyz$,  we   get   the homogeneous description
  of the elliptic fiber
\be
f_0(x,y,z)=x^2y+xy^2+ e^{-t_1}yz^2+e^{-t_2}xz^2+xyz. \ee
 This equation is a cubic polynomial in ${\bf P^2}$  whose  the  general
 form
    is given by
\be
\sum\limits_{i+j+k=3}a_{ijk}x^iy^jz^k=0, \ee
 which  can be written in the following Weierstrass form
\be
y^2z=x^3+axz^2+bz^3. \ee
 As we have seen,   this  form plays an  important role
  in the discussion of elliptic Calabi-Yau  manifolds  involved in
   F-theroy  to derive non perturbative vacua of type IIB
   superstring.
\subsection { K3 fibration   in F-theory compactifications}
\subsubsection{ LG $K3$ fibration  Calabi-Yau  superpotential}
Our second example of  Calabi-Yau  threefolds  is
  quite similar to the first one, and our treatment of it will parallel to
    the above discussion. This example  of model  will be  given by    the
       LG  mirror  superpotential with  a local toric Calabi-Yau  3-fold
 configuration, which is both elliptic and $K3$ fibration.
Roughly speaking,  the dual  field content of  $2D$ $N=2$  linear
sigma model  is  a   $ U(1)^5$  supersymmetric  gauge theory with
ten $(\phi_i)$ matter fields  with   vectors  charges $ q_i^a$ $
(a=0,\ldots,4) $.  The latters  are   the quantum charges of the
$(\phi_i)$'s under  the corresponding  $ U(1)^5$'s: \bea
q_i^0&=&(-2,0,1,0,0,1,0,0,0,0)\nn\\
q_i^1&=&(0,-2,1,0,0,0,0,1,0,0)\nn\\
q_i^2&=&(1,1,-2,1,1,0,0,0,0,-2)\\
q_i^3&=&(0,0,1,-2,0,0,0,1,0,0)\nn\\
q_i^4&=&(0,0,1,0,-2,0,0,0,0,1)\nn, \eea
 which are, up to some details, the opposite of the  affine $so(8)$
  Cartan
  matrix $K_{ai}(so(8))$:
\be
q^a_i= -K_{ai}(so(8)),\quad i=1,\ldots,5 ;\quad  a=0,\ldots,4 .
 \ee
The space  of classical vacua of this   model  is given by the
D-flatness equations namely
\be
\sum\limits _{i=1}^{10} q_i^a |\phi_i|^2=\rho_a,\quad
a=0,\ldots,4, \ee
This space of solutions has  also  a geometrical
realization described by the following toric data
\be
\sum\limits _{i=1}^{10} q_i^a { v_i}=0,\quad a=0,\ldots,4
\ee
 where
\bea { v_1}=(1,1,-1,-1), { v_2}=(1,-1,-1,1), {
v_3}=(2,-1,-1,-1),\nn\\ { v_4}=(1,-1,-1,-1), { v_5}=(1,-1,1,-1), {
v_6}=(0,3,-1,-1),\\ { v_7}=(0,-1,-1,3), {v_8}=(0,-1,-1,-1), {
v_9}=(0,-1,3,-1),\nn\\ { v_{10}}=(0,0,0,0).\nn \eea Using
equations (3.13-14) and recalling the variables,  the mirror
theory has superpotential
\be
W_3(x_i)=\sum\limits_{i=1}^{10}x_i=0, \ee  where the $x_i$'s
satisfy the following  constraint equations, \bea
x_3^2x^2_{10}&=&x_1x_2x_4x_5\nn\\ x_1^2&=&x_3x_6\nn\\
x_2^2&=&x_3x_7\\ x^2_4&=&x_3x_8\nn\\ x^2_5&=&x_3x_9.\nn \eea
 These constraints  can be  solved  by the monomials
\bea x_1=w z_1^2,\quad x_2=w z_2^2,\quad x_3=w ^2,\quad x_4=w
z_3^2,\quad  x_5=w z_4^2.\nn\\ x_6=z_1^4,\quad x_7=z_2^4,\quad
x_8=z_3^4,\quad x_9=z_4^4,\quad x_{10}=z_1z_2z_3z_4. \eea Thus,
the LG mirror superpotential is
\be
W_3=z_1^4+z_2^4+z_3^4+z_4^4+\psi z_1z_2z_3z_4+a_0w^2+w
(a_1z_1^2+a_2z_2^2+a_3z_3^2+a_4z_4^2)=0, \ee
 where $\psi$ and $a_i$, which given in terms of $t_i$, are complex
  parameters defining the complex structure.  \\
 Equation (4.28)  is invariant under the $C^*$ action
\be
  (z_1,z_2,z_3,z_4,w)\to (\l z_1,\l z_2,\l z_3,\l z_4,\l ^2w),
\ee
and describes  a 3-dimensional hypersurface in ${\bf WP^4}(1,1,1,1,2)$ with
 $c_1\neq 0$.  One easily restore the Calabi-Yau condition by considering
   $ w W_3$ as the Calabi-Yau  hypersurface which defines  a singular
    3-dimensional toric manifold.  This geometry  is
       not  only elliptic but  also    K3  fibration.  To  see this,
        consider  first the $w $ independent terms namely
\be
P_{\psi}=z_1^4+z_2^4+z_3^4+z_4^4+\psi z_1z_2z_3z_4.
\ee
 This defines  a  quartic  hypersurface in  ${\bf P^3}$
  describing a  $K3$  surface   with  a  complex  structure $\psi $.
   Second,  we take  the  large complex structure  limit  ($\psi\to \infty$).
   In this appropriate limit, the equation  (4.30) becomes
   approximately
\be
P_{\infty}=z_1z_2z_3z_4=0. \ee
 According to \cite{lv}, this means that   the quartic $K3$  is   now a  $T^2$
    fibration  over the boundary faces of the toric diagram $\Delta$
    of the
      ${\bf P}^3$   projective  space:
\be
K3=T^2(R_1,R_2)\t B_2,
\ee
where $(R_1,R_2)$ are the two  radii of the torus  $T^2$ and $B_2=\p \Delta$
  consists of  the union of  four    triangles of  three dimensional
   standard
   simplex.  Note that this   torus  can degenerate over the fixed faces of
   the $ \bf P^3$
    toric  action. One distinguishes two cases:\\
1- The torus degenerates to a circle at each edge, which  means
that one 1-cycle shrinks to zero size,
\be
R_i=0,\quad R_{j\neq i}\neq 0,\quad i,j=1,2.
\ee
This is the same situation appearing in the large complex structure limit
  of  elliptic curves involved  in   the study of   non perturbative
  vacua of type
   IIB string
   from
   F-theory compactifications  on   elliptic fibration manifolds.\\
2- The torus $ T^2$  completely  degenerates over  the  endpoints
of each triangle,
 where the two 1-cycles of $T^2$  shrink   to  zero size:
\be
R_i=0,\quad i=1,2.
\ee
 In these  singular limits, one  may    take   the  complex  structure $\psi $   of   $K3$ as
\be
\psi \sim  {V(B_2)\o R_1R_2},
\ee
 where  $V(B_2)$  denotes  the  volume  of   the  base  space  $ B_2$. In this  way,
 $W_3$ geometry  may  be now  regarded as fibering elliptic  $K3$ surface   ( 4.30),  in which
    the  fiber has vanishing first Chern class (i.e) $c _1=0$,
       over a  base space  parameterized  by   $w$.  Our   $W_3$
         Calabi-Yau  geometry has the following  nice features:  \\
          (1) It extends the geometry  of the $W_2$, studied in section 3,    to
\be
\sum\limits _iw^iP_i(z_1,z_2,z_3,z_4)=0, \ee
 for an  elliptic Calabi-Yau 3-folds.
 In other words,  instead   of    having   a curve in   two  dimensional  projective
  spaces( as in  the elliptic $K3$ surface (3.22)), we  now   have   a   surface
     $$ P_0(z_1,z_2,z_3,z_4)=0$$    in    the  three   dimensional  projective space $\bf
     P^3$;
  where  the  $w$ exponents are exactly the Dynkin index of     affine  $so(8)$  Lie
  algebra.\\
(2)   $W_3$   gives  a  new    realization of   $so(8)$  Lie
algebra in  terms of  Calabi -Yau  3-folds.  This toric
realisation  is closed  related  to  the standard   tetravalent
geometry \cite{bs3},  which may  be viewed as higher geometry  of
trivalent and bivalent geometries  used  in  the context  of
geometric engineering  of QFT  in  four dimensions \cite{kmv}. The
latter is described by the following monomials
\be
1,z_1^2,z_2^2,z_3^2,z_4^2,z_1z_2z_3z_4,  \ee  where this geometry
may be used to extend the $T_{p,q,r}$ singularity to $T_{p,q,r,t}$
by considering four intersecting $ SU$ chains.
\\ (3) The complex structure determined by the complex parameters
$ \psi$ and $ a_i$  maight be used to define  a moduli space of
SO(8) bundle on quartic $K3$.
\subsubsection{ $D=6$ $N=1$   SO(8) gauge theory }
 Having introduced the geometric background space $W_3$, we will  now discuss the
  corresponding gauge theory if one consider the F-theory compactification.
     As well known that    F-theory on  $K3$  fibration Calabi-Yau  manifolds   are
       intimately related to $N=1$ string models.  In  particular,  six dimensional
        compactifications of   F-theory on    Calabi-Yau threefolds, where  these geometries
         encode  the  informations  about    physical data  of $N=1$ string theories including
           the enhanced gauge symmetries,  the  perturbative  matter  fields  and    the
non perturbative dynamics  corresponding   to  small   instanton
singularities  \cite {ws}. Roughly speaking, mimicking  the
analysis in \cite{m1,m2}, F-theory  on  singular  $W_3$  (4.27)
leads to a  pure $N=1$  Yang-Mills theory  in  six  dimensions.
The corresponding  gauge group, associated  to this   $W_3$
geometry, is given by  the  SO(8)  gauge group  determined  by the
intersection  matrix of the blown up  toric divisors  (4.20).
 Moreover  since  the
perturbative gauge   symmetries  in  heterotic superstring models
stems only  from  the classification of  singularities of   $
W_2(ADE)$ fiber space, studied  in section  2, this  SO(8) gauge
model may be related to non perturbative dynamics.
\section{ On  ADE     Calabi-Yau  fourfolds  superpotentials }
In this  section we want to extend the previous analysis to higher
dimensional elliptic  Calabi-Yau   geometries.  In particular, we
will consider    $(n+2)$-dimensional  elliptic Calabi-Yau's, where
they      will be    realized   as   $n$-dimensional elliptic
Calabi-Yau  manifolds  over  2 complex dimensional base space.
They  may  be viewed  as few  extensions of  non compact
Calabi-Yau 3-folds
 (4.1).   These geometries   may  be expressed  in  the following   form
\be
f(x_1,\ldots,x_{n+1})=uv. \ee
 In other words,  instead   of  having a Reimann surface as  in the case of
   Calabi-Yau 3-folds, we now have  a $n$-dimensional  Calabi-Yau fibers
\be
f(x_1,\ldots,x_{n+1})=0. \ee
 These extended  geometries   may     play   a  crucial   role  in   the
  understanding  of    the   lower  dimensional      non  perturbative
   superstring  theories.  \\
From  the F-theory compactification point of  view,   we  will
restrict  to
 a  particular case  corresponding   to  elliptic  Calabi-Yau 4-folds
\be
f(x_1,x_2,x_{3})=uv.
\ee
    Before discussing  the $2D$ $N=2$   sigma  model  construction of
      these manifolds,  it  useful to review some basic facts about the
       different constructions  of the Calabi-Yau 4-folds. The latters can  have
        realizations of many types: \\
1- The orbifold   $\bf C^4 \o  Z_4$:
\be
z_j\to i z_j ,\quad j=1,\ldots,4.\nn \ee
 2- The hyper-Kahler quotient in terms of   two dimensional  field theory with eight
  supercharges in presence of charged  hypermultiples. \\  3 -The  ADE  hypersurfaces
   in  $\bf C^5$ considered in  \cite{gvw} in the context of derivations of two dimensional
    superconformal field  theories  from singular limits  of type IIA  superstring
     compactifications.
 \\
 Here  we are interested  in elliptic ADE  4-folds hypersurfaces  having  elliptic  $K3$  toric
fibration, with ADE singularities,  over  2-dimensional base
spaces.   A priori there  are   different  ways one may follow to
give the  corresponding  $2D$ $N=2$  linear sigma model
construction. A   naive  way  to  do  this  is   to consider these
geometries as a  moduli  space of   two orthogonal  models
described by    $ 2D$ $N=2$ supersymmetric  field  theories.  In
this method,  it is possible to see  the   elliptic  ADE  $N=2$
linear sigma model, studied in section 3,  as   a fiber   and  the
other model  whose the  target space  is   a two  complex
dimensional space,  as  a base.   However   this  way  of  doing
may  bring extra parameters in   the  moduli  space of   ADE
Calabi-Yau fourfolds hypersurfaces.   A  tricky  method  to
overcome this problem  is to use   the previous   elliptic    ADE
($K3$)   $N=2$ linear sigma  model  with extra  chiral fields,
corresponding to the  two complex  dimensional  base  space of
Calabi-Yau  fourfolds. Roughly speaking, we consider   the
previous $U(1)^{r+1}$  linear sigma model  but with   $(r+8)$
chiral fields   $\phi_j$ $( j=1,\ldots,r+8)$  with matrix  charge
$Q^a_j$. The latters are given by
\be
Q^a_j= (q_i^a, q_{r+6}^a,q_{r+7}^a,q_{r+8}^a ),\quad
i=1,\ldots,r+5,\;\; a=1,\ldots,r+1, \ee
 where $ q_i^a$ are exactly the matrix  charge of $ (r+5)$ chiral fields  $\phi_i$ of
     $U(1)^{r+1}$  linear sigma model
construction of ADE elliptic $K3$  and   $(q_{r+6}^a,q_{r+7}^a,q_{r+8}^a )$
  are the quantum charges of the extra fields,   under    $U(1)^{r+1}$
   symmetry,  will be specified  latter on.  The condition under which
    the gauge system  flow in the infrared  to $2D$ $N=2$ superconformal
    field theory is
\be
\sum\limits_{j=1}^{r+8}Q_j^a=0,\quad a=0,1,\ldots,r. \ee
 However  the Calabi-Yau condition $
\sum\limits_{i=1}^{r+5}q_i^a=0$   for   the  ADE elliptic $K3$
requires that
\be
q_{r+6}^a+q_{r+7}^a+q_{r+8}^a=0,\quad \forall a. \ee
 The vacuum energy of this  $N=2$ sigma  model is given by the D-flatness equations
\be
\sum\limits_{j=1}^{r+8}Q_j^a|\phi_j|^2= \rho^a,\quad a=0,\ldots,r,
\ee where  this  space of solutions, up to the identifications
imposed by the action of gauge group,  has  a toric realization.
This represented by $(r+8)$ vertices  $ V_j$ $(j=1,\ldots, r+8) $
of the standard lattice $\bf Z^5$ satisfying  the following toric
relations:
\be
\sum\limits_{j=1}^{r+8}Q_j^a  V_j= 0, \quad a=0,\ldots,r. \ee
Using the conventional notation $
V_j=V_{j\ell},\;\ell=1,\ldots,5$, the above  toric data ( 5.9) may
be split as \bea \sum\limits_{i=1}^{r+5}q_i^a  V_{i\ell'}&=
&0,\quad \ell'=1 ,2,3\\ q^a_{r+6}V_{r+6
\ell'}+q^a_{r+7}V_{r+7\ell'}+q^a_{r+8}V_{r+8\ell'}&=&0,\quad
\ell'=1 ,2,3 \\ \sum\limits_{j=1}^{r+8}Q_j^a  V_{j4}&=& 0,\\
\sum\limits_{j=1}^{r+8}Q_j^a  V_{j5}&=& 0. \eea  Equation (5.10)
is nothing  but the    equation   (3.18 )  where
\be
 V_{i\ell'}=v_i=(n_i,m_i,s_i),\quad \ell'=1,2,3,
\ee To write down  the LG mirror  superpotential,   we follow the
same analysis used in  section 3. This is obtained  in terms of
new gauge invariant monomials:
\be
x^j=\prod\limits_{\ell=1}^5x_{\ell}^{V_{j\ell}}. \ee Thus the ADE
mirror superpotentials are
\be
\sum \limits_{j=1}^{r+8}a_j\prod\limits_{\ell=1}^5x_{\ell}^{V_{j\ell}}=0.
\ee
 However  to work out the explicit form of  this  equation,
   we have to solve the toric  constraint equations (5.10-13).
      A  solution of  these  toric data   is  given
\bea V_i&=&(n_i,m_i,s_i,0,0),\quad i=1,\ldots, r+5\\
V_{r+6}&=&(0,0,0,\a,\a')\\ V_{r+7}&=&(0,0,0,\beta,\beta'),\\
V_{r+8}&=&(0,0,0,\gamma,\gamma') \eea where $ \a,\beta,\gamma,
\a',\beta'$  and $\gamma' $ are six   integers satisfying \bea
q_{r+6}^a\a+q_{r+7}^a \beta+q_{r+8}^a\gamma&=&0,\\
q_{r+6}^a\a'+q_{r+7}^a\beta'+q_{r+8}^a\gamma'&=&0. \eea
  For  latter use,    we choose a   special case where
\be
(q_{r+6}^a,q_{r+7}^a,q_{r+8}^a)=(1,-2,1),\quad   \forall  a. \ee
 In this way,   a  naive solution of equation (5.21-22)  is  given   by
 \be
( \a,\beta,\gamma)= (1,1,1),
\ee
  and
  \be
( \a',\beta', \gamma')=(-1,0,1).\ee
 Taking this special case, we get  the following  LG mirror   superpotential
\be
\sum\limits _{i=1}^{r+5} a_i {x_1}^{n_i}{x_2}^{m_i}{x_3}^{s_i}+x_4
({a_{r+6}\o x_5}+ a_{r+7}+ a_{r+8}x_5)=0. \ee However  using
equations (5.8) and (5.23), the mirror map  for $q_i^a=0\;
(i=1,\ldots r+5$), breaking the $U(1)^{r+1}$ symmetry to $U(1)$,
implies that the  LG fields corresponding to the mirror  base
geometry are constrained by
\be
{a_{r+6}\o x_5}+ a_{r+7}+ a_{r+8}x_5=0. \ee This means that the
mirror base geometry is zero-dimensional space.  From this
requirement, the mirror geometry of ADE hypersurfaces, obtained
after introducing the non dynamical fields, are given by
 \be
\sum\limits _{i=1}^{r+5} a_i {x_1}^{n_i}{x_2}^{m_i}{x_3}^{s_i}=uv
\ee where $ \sum\limits _{i=1}^{r+5} a_i
{x_1}^{n_i}{x_2}^{m_i}{x_3}^{s_i}=0$ is the equation of  the  ADE
elliptic K3 surfaces.  Finally if we consider F-theory
compactifications  on these elliptic hypersurfaces, we obtain
$D=4$  $N=1$ ADE gauge theories with non matter.
\section{Conclusion}
 In this paper,  we have studied  the Landau Ginzburg  theory  mirror to
$2D$  $ N=2$  gauged  linear  toric   sigma model.  We have
derived new classes  for  elliptic Calabi-Yau  superpotentials  of
Landau Ginzburg  theories.   The latters  play a crucial  role  
 in string/ brane physics. In the  Calabi-Yau  threefolds case,   we have
considered two examples of   the  mirror symmetry for toric sigma
model. First, we  have  given   the mirror theory of linear sigma
model on the canonical  line  bundle  over   the Hirzebruch
surfaces $\bf F_n$, recovering  the leading example of $ \bf F_0$
studied in the context of   the mirror action of Lagrangian
D-branes \cite{av}. In  this  case,   we  have shown that  the
mirror geometry   is an elliptic   Calabi-Yau threefolds whose the
fiber is   a Reimann surface.    Second we have found  a special
elliptic  and $K3$ fibration Calabi-Yau threefolds extending the
elliptic $K3$, considered in  the geometric engineering of  $4D$ $
N=2$ superconformal field theory, to   an  elliptic Calabi-Yau
threefolds with   affine so(8) Lie algebra Mori vectors. Moreover,
this geometry gives a new $N=1$    SO(8) pure Yang-Mills theory in
six dimensions from the F-theory compactification which may be
associated to non perturbative gauge symmetry in the  heterotic
string picture \cite{ws}. Finally,  we have  used the interplay
between toric geometry and gauged  linear sigma model to derive an
intuitive algebraic realization for the mirror superpotentials
associated to ADE Calabi-Yau fourfolds hypersurfaces.   \\ \\
 {\bf Acknowledgments}\\  Adil   Belhaj
would like to  thank the organizers of the Spring Workshop on
  Superstrings and related Matters (2001), the Abdus Salam International Centre for
    Theoretical Physics (ICTP), Trieste, Italy, for hospitality.  He  would
    like to
      thank E. H. Saidi  for  valuable  discussions,  encouragement and scientific helps.  He would  like also  to thank
      C. Gomez   and P. Mayr  for useful discussions, encouragement and scientific helps. He is grateful to
         R. Gopakumar,  D. Kutasov and  S.
        Theisen for
 discussions at ICTP. \\ This work is supported by SARS,
programme de soutien \`a la recherche scientifique; Universit\'e
Mohammed V-Agdal, Rabat.


\newpage



\begin{thebibliography}{99}

\bibitem{syz} Andrew Strominger, Shing-Tung Yau, Eric Zaslow,
 Mirror Symmetry is T-Duality;  Nucl. Phys. {\bf B479} (1996) 243-259,
  hep-th/9606040.
\bibitem{fhsv} S. Ferarra, J.A. Harvey, A . Strominger, and C. Vafa,
Second-Quantized mirror Symmetry; Phys. Lett {\bf B361}
(1995)59-65,   hep-th/9505162.
\bibitem{st}
S. Theisen, Intoduction to Calabi-Yau manifolds, Spring  Workshop
  on Superstrings and Related Matters, (ICTP, Trieste, 2001).
\bibitem{kkv}
S. Katz, A. Klemm and C. Vafa; Nucl. Phys {\bf B497}(1997)
173-195.
\bibitem{kmv}
 S. Katz, P. Mayr and C. Vafa,  Mirror symmetry and exact solution of
 4d  N=2 gauge theories I; Adv. Theor.  Math. Phys {\bf 1}(1998)53.
 \bibitem{m}  P. Mayr, Geometric Construction of N=2 gauge theories, Spring  Workshop
  on Superstrings and Related Matters, (ICTP, Trieste, March 1999).
 \bibitem{bfs1} A.Belhaj,  A. EL Fallah and E.H. Saidi, On the affine $ D_4$ mirror  geometry;  Class. Quantum. Grav.{\bf 16} (1999)3297-3306.
\bibitem{bfs2} A.Belhaj,  A. EL Fallah and E.H. Saidi,   On non simply laced   mirror geometries in type II strings;   Class. Quantum. Grav.{\bf 17} (1999)515-532.
\bibitem{hv}Kentaro Hori, Cumrun Vafa,  Mirror Symmetry,
hep-th/0002222.
\bibitem{hiv}Kentaro Hori,  Amer Iqbal,  Cumrun Vafa,  D-Branes And Mirror Symmetry,
hep-th/0005247.
  \bibitem{av} M. Aganagic and  C. Vafa,  Mirror Symmetry,  D-branes and
   Counting Holomorphic Discs,  hep-th/0012041.\\
     M. Aganagic, C. Vafa,
Mirror Symmetry and a $G_2$ Flop, hep-th/0105225.\\
 M. Aganagic, A. Klemm, C. Vafa, Disk Instantons,
Mirror Symmetry and the Duality Web, hep-th/0105045.
 \bibitem{vf}
 C. Vafa; Nucl.  Phys. B 469 (1996)403.
 \bibitem{vm} C. Vafa and D. Morrison; Nucl . Phys.  {\bf B 473} (1996)74; Nucl . Phys. B 476 (1996)437.
 \bibitem{gvw}
S. Gukov,  C. Vafa and E.  Witten, CFT'S From Calabi-Yau
Four-folds; Nucl.Phys. B584 (2000) 69-108; Erratum-ibid. B608
(2001) 477-478, hep-th/9906070.
 \bibitem{p}
J. Polchinski,  Phys, Rev. Lett {\bf 75} (1995)4724.
\bibitem{hw} A. Hanany and E. Witten; Nucl. Phys  {\bf B492}  (1997)152-190,  hep-th/9611230.
 \bibitem{wbrane}  E. Witten; Nucl. Phys  {\bf B500}  (1997)3-42, hep-th/9703166.
\bibitem{b}
A. Belhaj, On Geometric Engineering of Supersymmetric Gauge
Theories, the proceedings of the Workshop on Noncommutative
Geometry, Superstrings and Particle Physics. Rabat -Morocco,
(16-17 June 2000), hep-ph/0006248.
 \bibitem{sch}
  John H. Schwarz,  An  SL(2,Z)  Multiplet of Type IIB Superstrings,
Phys.Lett. {\bf  B360}  (1995) 13-18, hep-th/9508143.
 \bibitem{vw}
 C.Vafa and E. Witten; Nucl. Phys.Proc.Suppl {\bf 46}  (1996)225.
\bibitem{bm} P. Berglund and P. Mayr, Heterotic String/F-theory Duality
 from Mirror Symmetry; Adv.Theor.Math.Phys. 2 (1999) 1307-1372

            hep-th/9811217.
\bibitem{m1}  P. Mayr, Non -Perturbative N=1 String Vacua,
lectures delivered at the Spring Workshop on superstrings and
related matters, ICTP Trieste (March 1999).
\bibitem{m2}  P. Mayr, N=1 Heterotic string Vacua from Mirror
 Symmetry; lectures presented at the Winter School on Vector Bundles,
 Mirror Symmetry, and Lagrangian Submanifolds, Harvard University,
 January 1999, hep-th /9904115.
\bibitem{ht}
C. M. Hull, P. K. Townsend,  Unity of Superstring Dualities;
Nucl.Phys.   {\bf B438} (1995) 109-137, hep-th/9410167.
\bibitem{wd}
 E. Witten, String Theory Dynamics In Various Dimensions; Nucl.Phys. {\bf  B443}  (1995) 85-126,  hep-th/9503124.
\bibitem{ketal}
S. Kachru, A. Klemm, W. Lerche, P. Mayr, C. Vafa, Nonperturbative
Results on the Point Particle Limit of N=2 Heterotic String
Compactifications; Nucl.Phys. {\bf  B459}  (1996) 537-558,
hep-th/9508155.
\bibitem{kv}
 S. Kachru, C. Vafa, Exact Results for N=2 Compactifications of
 Heterotic Strings; Nucl. Phys.  {\bf B450}  (1995) 69-89, hep-th/9505105.
\bibitem{asp}
 Paul S. Aspinwall, Enhanced Gauge Symmetries and K3 Surfaces;  Phys. Lett.  {\bf  B357}(1995) 329-334,  hep-th/9507012.
\bibitem{w}  E. Witten; Nucl  Phys {\bf B403} (1993)159-22, hep-th/9301042.
\bibitem{agm} P. Aspinwall, B.R.  Grenne and R. Morrison;  Nucl. Phys   B416 (1994) 414.
\bibitem{f}W. Fulton, Introduction to Toric varieties; Annals of Math. Studies, No .131, Princeton University  Press, 1993.
\bibitem{c}
D. Cox, the homogeneous coordinate Ring of a toric variety,
J.Alg.geom. {\bf 4} (1995)17.
\bibitem{lv} N.C. Leung and C. Vafa; Adv .Theo . Math. Phys {\bf 2}(1998) 91, hep-th/9711013.
\bibitem{cpr}  Philip Candelas, Eugene Perevalov and Govindan Rajesh, Toric Geometry and Enhanced Gauge Symmetry of F-Theory/ Heterotic Vacua , Nucl.Phys. {\bf  B507} (1997) 445-474, hep-th/9704097.
  \bibitem{r}   Govindan Rajesh, Toric Geometry and F-Theory/ Heterotic Vacua Duality in Four Dimensions,  hep-th/9811240.
\bibitem{bs3}  A. Belhaj and  E.H.Saidi, Toric geometry, enhanced non
simply laced  gauge symmetries in superstring and F-theory
compactification,  hep-th/0012131.
\bibitem{bs1}   A. Belhaj and E. H. Saidi, Hyper-Kahler Singularities in
 Superstrings Compactification and 2d N=4 Conformal Field Theory; Class.Quant.Grav. 18 (2001) 57-82,  hep-th/0002205.
\bibitem{bs2}  A. Belhaj, E.H.Saidi, On HyperKahler Singularities;
Mod. Phys. Lett. A, Vol. 15, No.  {\bf 29}  (2000) pp. 1767-1779,
 hep-th/0007143.
\bibitem{v1}  C. Vafa, On N =1 Yang-Mills in Four Dimensions; Adv theor Math. Phys {\bf 2}
(1998).
\bibitem{swg1}  N. Seiberg and E. Witten; Nucl. Phys {\bf B426}(1994)19,  hep-th/9407087.
\bibitem{swg2}
 N. Seiberg and E. Witten;  Nucl. Phys {\bf B431}(1994)498,  hep-th/9408099.
\bibitem{ws}
 E. Witten, Small Instantons in String Theoy; Nucl.Pys. {\bf  B480} (1996)213, hep-th/9511030.
\end{thebibliography}
\end{document}